# Solutions for 80 km DWDM Systems

Annika Dochhan[1], Helmut Griesser[2], Michael Eiselt[1], and Jörg-Peter Elbers[2]
*1 ADVA Optical Networking SE, Maerzenquelle 1-3, 98617 Meiningen, Germany*
*2 ADVA Optical Networking SE, Fraunhoferstr. 9a, 82152 Martinsried, Germany*
ADochhan@advaoptical.com

**Abstract:** We review currently discussed solutions for 80 km DWDM transmission targeting inter data-center connections at 100G and 400G line rates. PDM-64QAM, PAM4 and DMT are investigated, while the focus lies on directly detected solutions.

## 1. Introduction

In today's internet environment with cloud services, video applications and other high-capacity storage and high speed applications, data center interconnects with a few tens of km reach are becoming an important market segment. The vast amount of data to be transmitted requires the use of dense WDM technology, but the short distance imposes strict limits on the acceptable costs per bit. Short distances relax the requirements on the optical signal-to-noise ratio (OSNR), so modulation formats with lower noise tolerance, such as directly detected (DD) solutions come into play. In this paper, we give a short overview of current 100G solutions and show future 400G candidates, namely discrete multi-tone transmission (DMT), 4-level pulse amplitude modulation (PAM4), both with DD, and polarization multiplexed 64-level quadrature amplitude modulation (PDM-64QAM) with coherent RX.

## 2. Requirements for 80 km DWDM transmission

In long-haul transmission systems, the most important quality criterions are the tolerance towards optical noise and the non-linearity of the transmission fiber. If several thousands of km need to be bridged, the attainable receiver OSNR is limited by the number of optical amplifiers in the system. Commercially available solutions for this target reach are PDM quadrature phase shift keying (PDM-QPSK) for 100 Gb/s [1], or DPSK (differentially detected binary PSK) at 40 Gb/s [2]. However, if the transmission distance is shorter and therefore less noise is accumulated, several other modulation formats with much higher OSNR requirement come into play. This includes the use of multi-level signals to transport high data rates at low bandwidth occupation. These formats require high receiver sensitivity, so at least an optical pre-amplifier is necessary for an 80 km system. For a fully loaded 96-channel DWDM system, the use of erbium doped fiber amplifiers (EDFAs) and thus transmission in the 1550-nm-wavelength window is advisable. Besides this, stringent requirements on cost, power consumption and footprint apply and an implementation with arrayed transmitters on photonic integrated circuits (PICs) is desirable [3].

## 3. From 100G to 400G

Today's commercial 100G short reach solutions divide into coherent DP-QPSK systems at ~25 GBaud with relaxed specifications, e.g. in terms of DSP capabilities and cheaper optics, and into directly detected binary NRZ or optical duobinary (ODB) systems with ~25 Gb/s. DD systems require 4 wavelengths to transmit 100 Gb/s per second but outperform the coherent solution in terms of cost and power efficiency. These solutions enable transmission over several hundreds of km, which is more than the targeted reach between the sites of a data center. Moreover, the huge amount of data traffic needs an upgrade in spectral efficiency and thus cannot be handled only by adding more wavelengths – other solutions for 80 km and 400G have to be found.

On the base of 100G low cost coherent solutions one might consider a coherent solution with 400G per wavelength. The cost of broadband DSP and other electrical components and the allowance of high OSNR requirements lead to the suggestion of PDM-64-QAM at a symbol rate of ~33.3 GBaud (plus overhead). Recent experiments have demonstrated a reach of up to 160 km of SSMF (300km ULAF) [4] and 600 km of EX3000/EX2000 fiber [5], running at 43 GBaud. In both cases, strong forward error correction codes (FEC) are necessary to enable transmission, leading to overheads of ~30%. These experiments were all carried out in lab environments with high quality components and the route to solutions with inexpensive, small footprint integrated circuits may be long. Moreover, compared to DD solutions, the realization of the coherent receiver DSP blocks, including the high-performance FEC, might be extremely challenging if the power consumption is restricted. Therefore, for the rest of the paper we focus on DD solutions, namely PAM4 and DMT. In this case, multi-wavelength solutions are envisioned, e.g. a superchannel comprising 8 wavelengths carrying 56 Gb/s each or a flexible scheme as shown for DMT below, where 4 to 8 wavelengths are chosen depending on the desired reach.

Since the DMT signal is constructed out of numerous low frequency subcarriers which can be loaded with individual modulation formats (bit and power loading: PL, BL), it can effectively compensate for any channel impairment, such as bandwidth limitations of components and power fading after DD due to chromatic dispersion (CD) [6]. Thus, single wavelength transmission of 112 Gb/s over 10 km of SSMF in C-Band was successfully demonstrated [7, 8], but at longer distances power fading significantly reduces the available bandwidth. Most recent research focusses on the application of DMT for O-Band client side optics [9, 10]. Here, [11] gives a comparison of DMT with PAM4 and PAM8 for 112 Gb/s single wavelength transmission with DMT showing superior performance.

Figure 1 (a) shows theoretical back-to-back (b2b) OSNR requirements for PAM4 at 56 GBaud (assuming an optical matched filter receiver), DMT at 56 GBaud using either QPSK or 16QAM on all subcarriers and PDM-64QAM at 448 Gb/s. In practical DMT systems the bandwidth limitations due to DAC and ADC roll-off and from other components can effectively be compensated by BL and PL. In order to show the theoretical limits, however, for Fig.1 we used rectangular channel filters with no roll-off to maintain the available bandwidth for the signals. In fact, the 16QAM DMT signal requires half the bandwidth of the QPSK DMT signal. For 64QAM we assume 448 Gb/s, but as predicted by the measurements mentioned above, the need for FEC might enhance the required line rate.

The performance of PAM4 significantly depends on the quality of the transmit signal and the applied pre- and post-equalization. E. g. [12] gives an overview on the influence of the system parameters for PAM4 and PAM8 using a Silicon Photonic modulator in 1.3 μm wavelength range and transmission of up to 20 km SSMF. The use of high resolution DACs for signal generation enables pre-processing of the signal, but also limits the bandwidth, so it might be advantageous to simply combine large bandwidth binary electrical signals without pre-equalization [13-17]. In addition to PAM4 at 1.3 μm [12, 15] and 850 nm [18], transmission in the targeted 1.5 μm window has been demonstrated using VCSELs [13, 14, 16] and EMLs [17]. With the VCSEL transmitter, 100 Gb/s PAM3 and PAM4 over several hundreds of km of fiber have been shown, but PDM and a coherent RX were included. [17] targeted at passive optical networks and transmitted 4x25 Gb/s PAM4 over up to 30 km using commercial 10Gb/s TOSA and ROSA components. For higher rate applications without coherent RX, CD becomes a severe limitation for PAM4.

Figure 1 also shows the simulated CD tolerance of PAM4 and DMT with electrical Bessel filters at TX (15 GHz) and RX (18 GHz). For PAM4 the performance depends mainly on the choice of amplitude levels [19], therefore two important special cases, equidistant levels of the optical field and equidistant levels of the optical power were considered. In addition, the use of a 13-tap symbol-spaced feed forward equalizer for improving the CD tolerance is added. The fiber is SSMF with dispersion coefficient 17 ps/nm/km [20]. It can be seen that although PAM4 has better tolerance for low distances, the transmission reach of DMT with BL and PL is significantly higher.

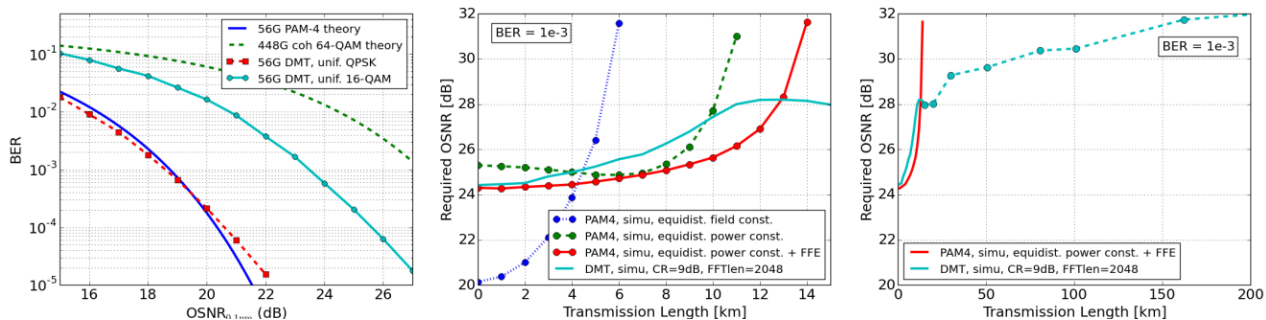

Fig. 1: Theoretical b2b OSNR performance for PAM4, DMT and PDM-64QAM (a), simulated transmission reach for PAM4 and DMT.

Figure 2 shows the measured OSNR performance of DMT for different rates for b2b and after 82 km dispersion uncompensated transmission over SSMF in the 1.5 μm window. A detailed description of the system and the DSP are given in [21] and [22]. BL and PL are performed according to the estimated SNR, which is determined by probing the channel with uniform BL and PL on all subcarriers. The SNR for 82 km and BL for 56 Gb/s are also displayed. If a FEC limit of 4e-3 is assumed, 112 Gb/s on a single wavelength is possible b2b, whereas the data rate after 82 km is limited to 76 Gb/s unless Raman amplification or other methods for increasing the OSNR performance are used. However, if SD-FEC is assumed with a limit of 1.9e-2, 96 Gb/s could be transmitted, although a higher FEC overhead would reduce the net data rate. Moreover, optical CD compensation can help to mitigate power fading. The results in Fig. 2 were obtained with broad optical filters, so no limitations on the optical signal bandwidth apply. However, to the keep spectral efficiency high, a 50 GHz WDM grid is desirable for multiple-wavelengths solutions. In [23] we showed the impact of narrow band optical filtering on the double sideband (DSB) DMT signal and proposed an asymmetrical vestigial sideband (VSB) approach. Recently [24], we enhanced these results and showed a flexible 4 to 8 wavelengths system, carrying 448 Gb/s. Figure 3 shows the reach of a single channel in case of symmetrical and asymmetrical filtering at different data rates, required to form 448 Gb/s in a WDM system. Filtering is performed by detuning the laser frequency, thus using the edge of the interleaver (IL) filter to generate the VSB signal. The filtering principle and the estimated SNRs for DSB and VSB after 80 km transmission are also shown. It should be mentioned that tolerance of laser detuning is in the order of 5 GHz. Finally, the BERs of all wavelength channels are displayed for the maximum reach of each channel count/data rate combination. Due to linear cross talk, reaches for WDM are smaller than indicated by the single channel results.

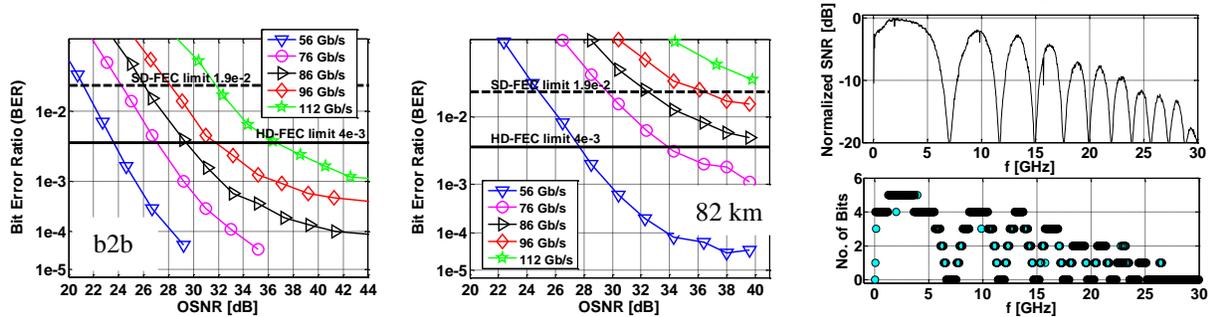

Fig. 2: BER vs. OSNR for DMT for different data rates: b2b, after 82 km transmission and estimated SNR and BL for 82 km and 56 Gb/s.

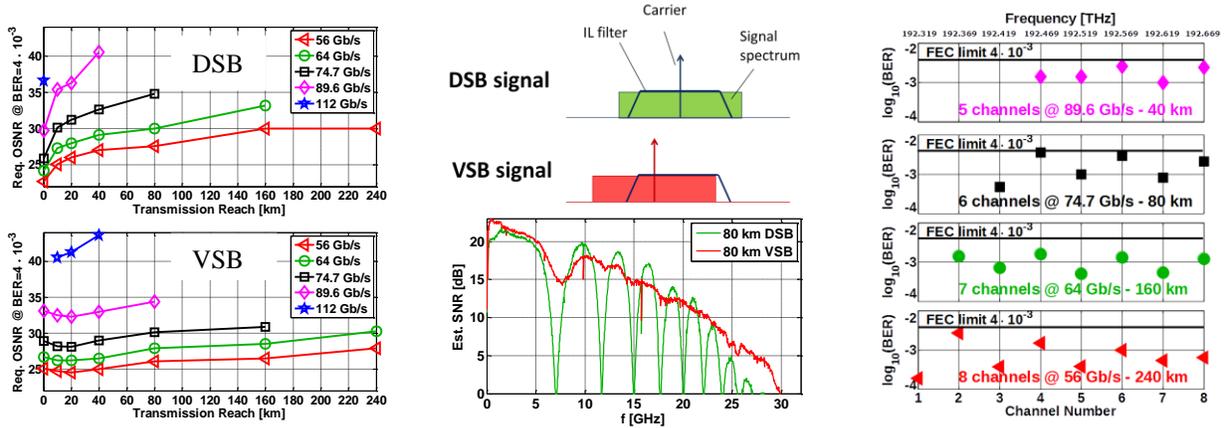

Fig. 3: Flexible 400G transmitter: Required OSNR vs. reach for various single channel data rates, DSB and VSB, principle of VSB and estimated SNR after 80 km for DSB and VSB, BER after transmission in 400G WDM system for different rates, no. of channels and reaches.

## 5. Conclusion
We reviewed several options for next generation 400G 80 km inter datacenter connections in the 1550 nm window. DD multi-level solutions promise low-cost and high integrability together with an increased spectral efficiency (c.t. 100G DD solutions). Especially DMT shows the potential to operate at high line rates without CD compensation, whereas PAM4 requires optical CD compensation in addition to electronic equalization. However, a fair comparison is only possible, if both solutions were evaluated in the same EDFA-amplified 1550 nm 80 km transmission environment. Raman amplification can help to increase the OSNR margin or allow higher rates per wavelength.

## 6. Acknowledgements

The results were obtained in the framework of the SASER-ADVAntage-NET project, partly funded by the German ministry of education and research (BMBF) under contract 16BP12400.